%% file: main.tex
\documentclass[times,twocolumn]{aastex62}
\usepackage{CJK}
\usepackage{amsmath}
\usepackage{multirow}
\usepackage{cases}
\usepackage{graphicx}
\usepackage{subfigure}
\usepackage{natbib}
\usepackage{color}
\usepackage{tabularx}
\usepackage{bm}
\usepackage{threeparttable}
\usepackage{gensymb}

\newcommand{\thetae}{\theta_{\rm E}}

\newcommand{\pie}{\pi_{\rm E}}

\newcommand{\te}{t_{\rm E}}
\newcommand{\event}{KMT-2025-BLG-1616}

\newcommand{\hjd}{{\rm HJD}^{\prime}}

\DeclareGraphicsExtensions{.pdf,.png,.jpg}

\shorttitle{}
\shortauthors{Yang et al.}

\begin{document}
\begin{CJK*}{UTF8}{gbsn}

\title{{\large KMT-2025-BLG-1616Lb: First Microlensing Bound  Planet From DREAMS}}

\correspondingauthor{Hongjing Yang, Weicheng Zang}
\email{hongjing.yang@qq.com, zangweicheng@westlake.edu.cn}

\input{author.tex}

\input{abstract}

\section{Introduction}\label{intro}
\input{intro}

\section{Observations}\label{obser}
\input{obser}

\section{Light-curve Analysis}\label{model}
\input{model}

\section{Source and Lens Properties}\label{lens}

\input{lens}

\section{Discussion}\label{dis}

\input{dis}

\bibliography{Zang.bib}

\end{CJK*}
\end{document}

%% file: author.tex
\author[0000-0003-0626-8465]{Hongjing Yang (杨弘靖)}
\affiliation{Westlake Institute for Advanced Study, Hangzhou 310030, Zhejiang Province, China}
\affiliation{Department of Astronomy, Westlake University, Hangzhou 310030, Zhejiang Province, China}

\author[0000-0001-6000-3463]{Weicheng Zang (臧伟呈)}
\affiliation{Center for Astrophysics $|$ Harvard \& Smithsonian, 60 Garden St., Cambridge, MA 02138, USA}
\affiliation{Department of Astronomy, Westlake University, Hangzhou 310030, Zhejiang Province, China}

\author[0000-0001-9823-2907]{Yoon-Hyun Ryu} 
\affiliation{Korea Astronomy and Space Science Institute, Daejon 34055, Republic of Korea}

\author{Takahiro Sumi}
\affiliation{Department of Earth and Space Science, Graduate School of Science, The University of Osaka, Toyonaka, Osaka 560-0043, Japan}

\author[0000-0002-1279-0666]{Jiyuan Zhang (张纪元)}
\affiliation{Department of Astronomy, Tsinghua University, Beijing 100084, China}

\author[0009-0001-6584-7187]{Hongyu Li (李弘禹)}
\affiliation{Department of Astronomy, Tsinghua University, Beijing 100084, China}

\author[0000-0002-2641-9964]{Cheongho Han}
\affiliation{Department of Physics, Chungbuk National University, Cheongju 28644, Republic of Korea}

\collaboration{(Leading Authors)}

\author[0000-0001-5651-9440]{Yuchen Tang}
\affiliation{Department of Astronomy, Westlake University, Hangzhou 310030, Zhejiang Province, China}

\author[0000-0003-4625-8595]{Qiyue Qian}
\affiliation{Department of Astronomy, Tsinghua University, Beijing 100084, China}

\author[0000-0002-1287-6064]{Zhixing Li}
\affiliation{Department of Astronomy, Tsinghua University, Beijing 100084, China}

\author[0009-0005-0410-8451]{Yuxin Shang}
\affiliation{Department of Astronomy, Tsinghua University, Beijing 100084, China}

\author[0000-0003-3201-061X]{Xikai Shan}
\affiliation{Department of Astronomy, Tsinghua University, Beijing 100084, China}

\author[0000-0001-8317-2788]{Shude Mao}
\affiliation{Department of Astronomy, Westlake University, Hangzhou 310030, Zhejiang Province, China}

\author{Guillermo Damke}
\affiliation{Cerro Tololo Inter-American Observatory/NSF’s NOIRLab, Casilla 603, La Serena, Chile}

\author{Alfredo Zenteno}
\affiliation{Cerro Tololo Inter-American Observatory/NSF’s NOIRLab, Casilla 603, La Serena, Chile}

\author{Steve Heathcote}
\affiliation{Cerro Tololo Inter-American Observatory/NSF’s NOIRLab, Casilla 603, La Serena, Chile}

\author{Konstantina Boutsia}
\affiliation{Cerro Tololo Inter-American Observatory/NSF’s NOIRLab, Casilla 603, La Serena, Chile}

\author[0000-0001-7016-1692]{Przemek Mr\'{o}z}
\affiliation{Astronomical Observatory, University of Warsaw, Al. Ujazdowskie 4, 00-478 Warszawa, Poland}

\author{Xiurui Zhao}
\affiliation{Cahill Center for Astrophysics, California Institute of Technology, 1216 East California Boulevard, Pasadena, 91125, CA, USA}

\author{Matthew Penny}
\affiliation{Department of Physics and Astronomy, Louisiana State University, Baton Rouge, LA 70803, USA}

\author{Sean Terry}
\affiliation{Code 667, NASA Goddard Space Flight Center, Greenbelt, MD 20771, USA}
\affiliation{Department of Astronomy, University of Maryland, College Park, MD 20742, USA}

\author{Patrick Tamburo}
\affiliation{Center for Astrophysics $|$ Harvard \& Smithsonian, 60 Garden St., Cambridge, MA 02138, USA}

\author{Timothy Cunningham}
\affiliation{Center for Astrophysics $|$ Harvard \& Smithsonian, 60 Garden St., Cambridge, MA 02138, USA}

\author{Quanzhi Ye}
\affiliation{Department of Astronomy, University of Maryland, College Park, MD 20742, USA}
\affiliation{Center for Space Physics, Boston University, 725 Commonwealth Ave, Boston, MA 02215, USA}

\author{Eric W. Peng}
\affiliation{NSF NOIRLab, 950 N. Cherry Avenue, Tucson, AZ 85719, USA}

\author{Rachel Street}
\affiliation{Las Cumbres Observatory Global Telescope Network, Inc., 6740 Cortona Drive, Suite 102, Goleta, CA 93117, USA}

\author{Katarzyna Kruszy{\'n}ska}
\affiliation{Las Cumbres Observatory Global Telescope Network, Inc., 6740 Cortona Drive, Suite 102, Goleta, CA 93117, USA}

\author{Etienne Bachelet}
\affiliation{IPAC, Mail Code 100-22, Caltech, 1200 E. California Blvd., Pasadena, CA 91125, USA}

\author{Yiannis Tsapras}
\affiliation{Astronomisches Rechen-Institut, M\"{o}nchhofstr. 12-14, D-69120 Heidelberg, Germany}

\author{Markus Hundertmark}
\affiliation{Astronomisches Rechen-Institut, M\"{o}nchhofstr. 12-14, D-69120 Heidelberg, Germany}

\collaboration{(The DREAMS Collaboration)}

\author[0000-0003-3316-4012]{Michael D. Albrow}
\affiliation{University of Canterbury, School of Physical and Chemical Sciences, Private Bag 4800, Christchurch 8020, New Zealand}

\author[0000-0001-6285-4528]{Sun-Ju Chung}
\affiliation{Korea Astronomy and Space Science Institute, Daejeon 34055, Republic of Korea}

\author{Andrew Gould} 
\affiliation{Max-Planck-Institute for Astronomy, K\"onigstuhl 17, 69117 Heidelberg, Germany}
\affiliation{Department of Astronomy, Ohio State University, 140 W. 18th Ave., Columbus, OH 43210, USA}

\author[0000-0002-9241-4117]{Kyu-Ha Hwang}
\affiliation{Korea Astronomy and Space Science Institute, Daejeon 34055, Republic of Korea}

\author[0000-0002-0314-6000]{Youn Kil Jung}
\affiliation{Korea Astronomy and Space Science Institute, Daejeon 34055, Republic of Korea}
\affiliation{National University of Science and Technology (UST), Daejeon 34113, Republic of Korea}

\author[0000-0002-4355-9838]{In-Gu Shin}
\affiliation{Department of Astronomy, Westlake University, Hangzhou 310030, Zhejiang Province, China}


\author[0000-0003-1525-5041]{Yossi Shvartzvald}
\affiliation{Department of Particle Physics and Astrophysics, Weizmann Institute of Science, Rehovot 7610001, Israel}

\author[0000-0001-9481-7123]{Jennifer C. Yee}
\affiliation{Center for Astrophysics $|$ Harvard \& Smithsonian, 60 Garden St., Cambridge, MA 02138, USA}

\author{Dong-Jin Kim}
\affiliation{Korea Astronomy and Space Science Institute, Daejeon 34055, Republic of Korea}

\author[0000-0003-0043-3925]{Chung-Uk Lee}
\affiliation{Korea Astronomy and Space Science Institute, Daejeon 34055, Republic of Korea}

\author[0000-0002-6982-7722]{Byeong-Gon Park}
\affiliation{Korea Astronomy and Space Science Institute, Daejeon 34055, Republic of Korea}

\collaboration{(The KMTNet Collaboration)}

\author{David P.~Bennett}
\affiliation{Code 667, NASA Goddard Space Flight Center, Greenbelt, MD 20771, USA}
\affiliation{Department of Astronomy, University of Maryland, College Park, MD 20742, USA}

\author{Ian A. Bond}
\affiliation{School of Mathematical and Computational Sciences, Massey University, Auckland 0745, New Zealand}

\author{Giuseppe Cataldo}
\affiliation{NASA Goddard Space Flight Center, Greenbelt, MD 20771, USA}

\author{Ryusei Hamada}
\affiliation{Department of Earth and Space Science, Graduate School of Science, The University of Osaka, Toyonaka, Osaka 560-0043, Japan}

\author{Yuki Hirao}
\affiliation{Institute of Astronomy, Graduate School of Science, The University of Tokyo, 2-21-1 Osawa, Mitaka, Tokyo 181-0015, Japan}

\author{Asahi Idei}
\affiliation{Department of Earth and Space Science, Graduate School of Science, The University of Osaka, Toyonaka, Osaka 560-0043, Japan}

\author{Shuma Makida}
\affiliation{Department of Earth and Space Science, Graduate School of Science, The University of Osaka, Toyonaka, Osaka 560-0043, Japan}

\author{Shota Miyazaki}
\affiliation{Institute of Space and Astronautical Science, Japan Aerospace Exploration Agency, 3-1-1 Yoshinodai, Chuo, Sagamihara, Kanagawa 252-5210, Japan}

\author{Tutumi Nagai}
\affiliation{Department of Earth and Space Science, Graduate School of Science, The University of Osaka, Toyonaka, Osaka 560-0043, Japan}

\author{Togo Nagano}
\affiliation{Department of Earth and Space Science, Graduate School of Science, The University of Osaka, Toyonaka, Osaka 560-0043, Japan}

\author{Seiya Nakayama}
\affiliation{Department of Earth and Space Science, Graduate School of Science, The University of Osaka, Toyonaka, Osaka 560-0043, Japan}

\author{Mayu Nishio}
\affiliation{Department of Earth and Space Science, Graduate School of Science, The University of Osaka, Toyonaka, Osaka 560-0043, Japan}

\author{Kansuke Nunota}
\affiliation{Department of Earth and Space Science, Graduate School of Science, The University of Osaka, Toyonaka, Osaka 560-0043, Japan}

\author{Ryo Ogawa}
\affiliation{Department of Earth and Space Science, Graduate School of Science, The University of Osaka, Toyonaka, Osaka 560-0043, Japan}

\author{Ryunosuke Oishi}
\affiliation{Department of Earth and Space Science, Graduate School of Science, The University of Osaka, Toyonaka, Osaka 560-0043, Japan}

\author{Yui Okumoto}
\affiliation{Department of Earth and Space Science, Graduate School of Science, The University of Osaka, Toyonaka, Osaka 560-0043, Japan}

\author[0000-0001-5069-319X]{Nicholas J. Rattenbury}
\affiliation{Department of Physics, University of Auckland, Private Bag 92019, Auckland, New Zealand}

\author[0000-0002-1228-4122]{Yuki K. Satoh}
\affiliation{College of Science and Engineering, Kanto Gakuin University, Yokohama, Kanagawa 236-8501, Japan}

\author{Daisuke Suzuki}
\affiliation{Department of Earth and Space Science, Graduate School of Science, The University of Osaka, Toyonaka, Osaka 560-0043, Japan}

\author[0000-0002-6510-0681]{Motohide Tamura}
\affiliation{Astrobiology Center, 2-21-1 Osawa, Mitaka-shi, Tokyo 181-8588, Japan}
\affiliation{Department of Astronomy, University of Tokyo, 7-3-1 Hongo, Bunkyo-ku, Tokyo 113-0033, Japan}

\author{Takuto Tamaoki}
\affiliation{Department of Earth and Space Science, Graduate School of Science, The University of Osaka, Toyonaka, Osaka 560-0043, Japan}

\author{Hibiki Yama}
\affiliation{Department of Earth and Space Science, Graduate School of Science, The University of Osaka, Toyonaka, Osaka 560-0043, Japan}

\collaboration{(The PRIME Collaboration)}

%% file: abstract.tex
\begin{abstract}
We present observations and analysis of the bound planetary microlensing event KMT-2025-BLG-1616. The planetary signal was captured by the Korea Microlensing Telescope Network (KMTNet) and the DECam Rogue Earths and Mars Survey (DREAMS). DREAMS's minute-cadence observations break the central/resonant degeneracy in the binary-lens models. The color of the faint source star ($I=22$) is measured from the DREAMS's $r - z$ color. The planetary system has a planet-host mass ratio of $q \sim 5 \times 10^{-4}$. A Bayesian analysis yields a host-star mass of $\sim 0.3\,M_\odot$, a planetary mass of $\sim 40\,M_{\oplus}$, a projected planet-host separation of $\sim 1.6~\mathrm{au}$, and a lens distance of $\sim 7.5~\mathrm{kpc}$. Based on the photometric precision achieved by DREAMS for this event, we simulate free-floating planet (FFP) detections and find that DREAMS is sensitive to Mars-mass FFPs in the Galactic bulge and Moon-mass FFPs in the Galactic disk.

\end{abstract}

%% file: intro.tex
The gravitational microlensing technique is sensitive to wide-orbit planets \citep{Shude1991,Andy1992} and to free-floating planets (FFPs) \citep{Sumi2011}. Two main challenges for microlensing planet searches are the low event rate and the short duration of planetary signals. To overcome the challenges, the first phase searches adopted the two-step strategy proposed by \citet{Andy1992}: large-area ($\sim100\ \mathrm{deg}^2$), low-cadence surveys $\Gamma \lesssim 1\ \mathrm{day}^{-1}$ to identify microlensing events, with high-cadence follow-up to capture planetary perturbations \citep[e.g.,][]{OB050071,OB050071D,OB05390}. This approach produced two homogeneous samples of bound planets and indicated that Neptune-class planets on wide orbits are common \citep{mufun,Cassan2012}, but it is poorly suited to detecting FFPs.

With the deployment of wide-field cameras on 1--2 meter telescopes, the second phase of microlensing planetary search, including the second phase of Microlensing Observations in Astrophysics (MOA-II, 2006+, \citealt{Sako2008}), the fourth phase of the Optical Gravitational Lensing Experiment (OGLE-IV, 2011+, \citealt{OGLEIV}), the Wise microlensing survey (2011--2014, \citealt{Wise}), the Korea Microlensing Telescope Network (KMTNet, 2016+, \citealt{KMT2016}), the Prime Focus Infrared Microlensing Experiment (PRIME, 2024+, \citealt{PRIME}), has carried out large-area monitoring with cadences ranging from $\Gamma\sim1\ \mathrm{day}^{-1}$ to $\sim6\ \mathrm{hr}^{-1}$. These survey-only programs have discovered more than 200 bound planets \citep{NASAExo} and produced four statistical samples \citep{Wise,Suzuki2016,OGLE_wide,OB160007}. The largest of these samples \citep{OB160007} suggests two populations of wide-orbit planets: gas giants and super-Earths/mini-Neptunes. At the same time, pure survey searches have reported $>10$ candidate FFPs (e.g., \citealt{Mroz2017a,OB161928,Gould2022_FFP_EinsteinDesert,Naoki_FFP}). If confirmed, these detections would imply that terrestrial- and super-Earth-mass FFPs are several to dozens of times more numerous than bound planets or other stellar objects \citep{Mroz2017a,Gould2022_FFP_EinsteinDesert,Sumi2023}.

The FFP mass functions inferred by \citet{Gould2022_FFP_EinsteinDesert} and \citet{Sumi2023} favor a power-law index near $-1$. If this trend extends to Mars- and even Moon-mass scales, the implied occurrence rate would exceed that inferred from current second-generation microlensing surveys by one to two orders of magnitude. In this low-mass regime, finite-source effects dominate the light curves' morphology, and as a result, the event rate scales with the occurrence rate of FFPs and independent of the angular Einstein radius and thus the mass \citep{CMST}. Consequently, wide-area coverage ($\sim100~\mathrm{deg}^2$) is unnecessary and a small-area survey (a few square degrees) with sufficient photometric precision and high cadence can efficiently detect low-mass FFPs.

Based on this idea, the DECam Rogue Earths And Mars Survey (DREAMS)\footnote{\url{https://time-allocation.noirlab.edu/\#/proposal/details/560332}}, using the $3~\mathrm{deg}^2$ Dark Energy Camera (DECam; \citealt{DECam2008,DECam2015}) on the 4\,m Blanco telescope at Cerro Tololo Inter-American Observatory (CTIO) in Chile, has been surveying a $5~\mathrm{deg}^2$ field since June 2025. With the larger aperture and a cadence of one exposure every 1--2 minutes, DREAMS accumulates about 50-500 times more source-star flux over the same sky area than other microlensing surveys, yielding substantially higher sensitivity to low-mass FFPs.

Beyond its high sensitivity to low-mass FFPs, DREAMS can also recover weak planetary signals in bound systems that other microlensing surveys miss or cannot adequately characterize. Such signals can arise from very low planet-host mass ratios ($q$; to date, only one planet with $\log q<-5$ has been discovered; \citealt{OB160007}), from weak cusp crossings, or from high-magnification events that require higher cadence and photometric precision (e.g., KMT-2022-BLG-0440; \citealt{KB220440}).

Here we present the first bound planet reported by DREAMS, KMT-2025-BLG-1616Lb, identified during a high-magnification event. The structure of this paper is as follows. In Section~\ref{obser} we describe the KMTNet, PRIME, and DREAMS observations of this event. Section~\ref{model} presents the light-curve analysis, and Section~\ref{lens} details the source and lens properties. Finally, in Section~\ref{dis} we present results using KMTNet and PRIME data only and estimate DREAMS's detection limits for FFPs.

%% file: obser.tex
\begin{figure*}
    \centering
    \includegraphics[width=0.85\linewidth]{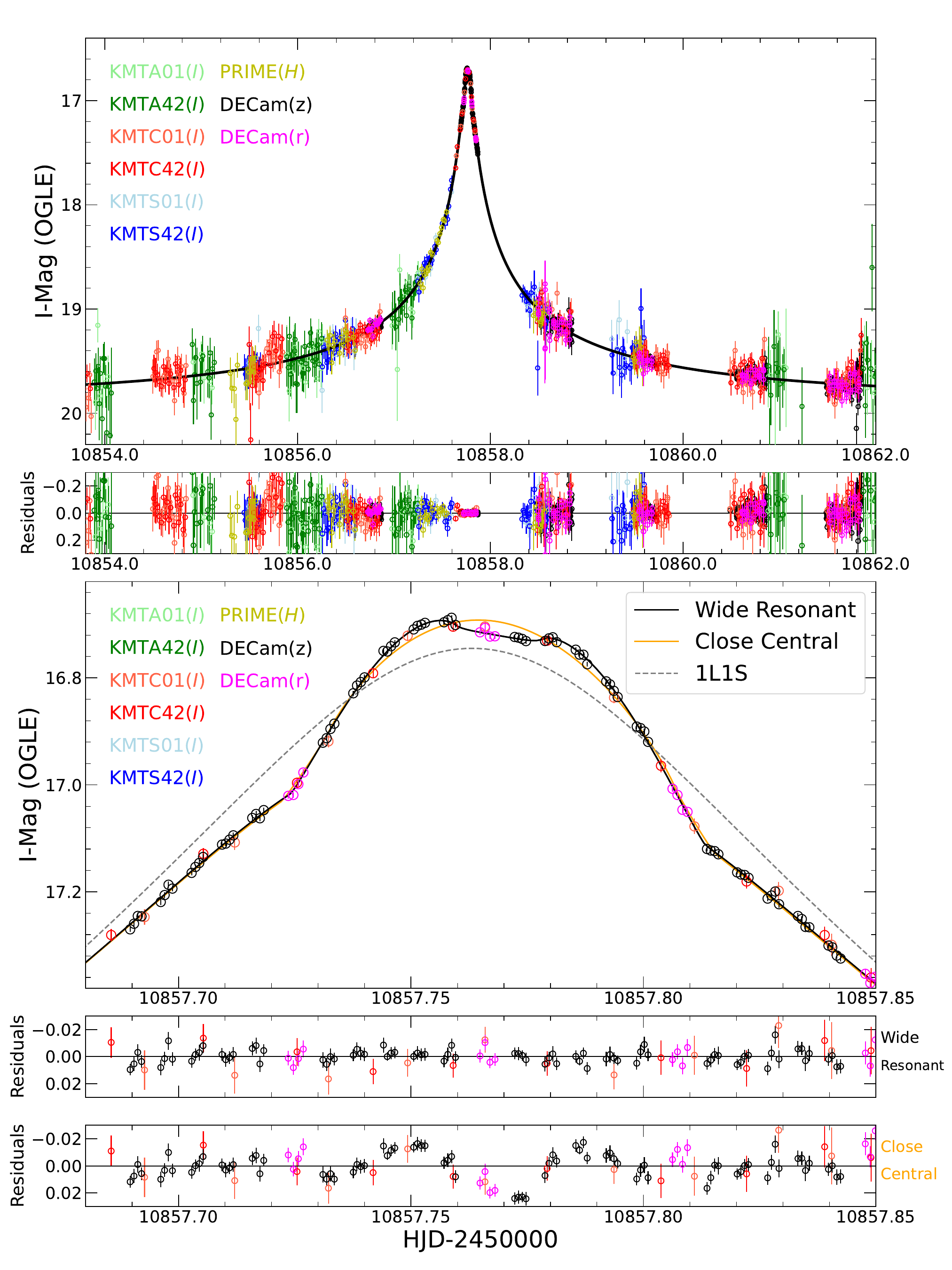}
    \caption{Light curve of the microlensing event, \event, with the 2L1S models (solid black and orange lines) and the underlying 1L1S model (dashed grey line). Different data sets are plotted in different colors. The upper panels shows the 8-day data around the peak. The lower panels present a close-up of the planetary anomaly and the residuals relative to the 2L1S models.}
    \label{fig:lc}
\end{figure*}

\begin{figure}
    \includegraphics[width=0.47\textwidth]{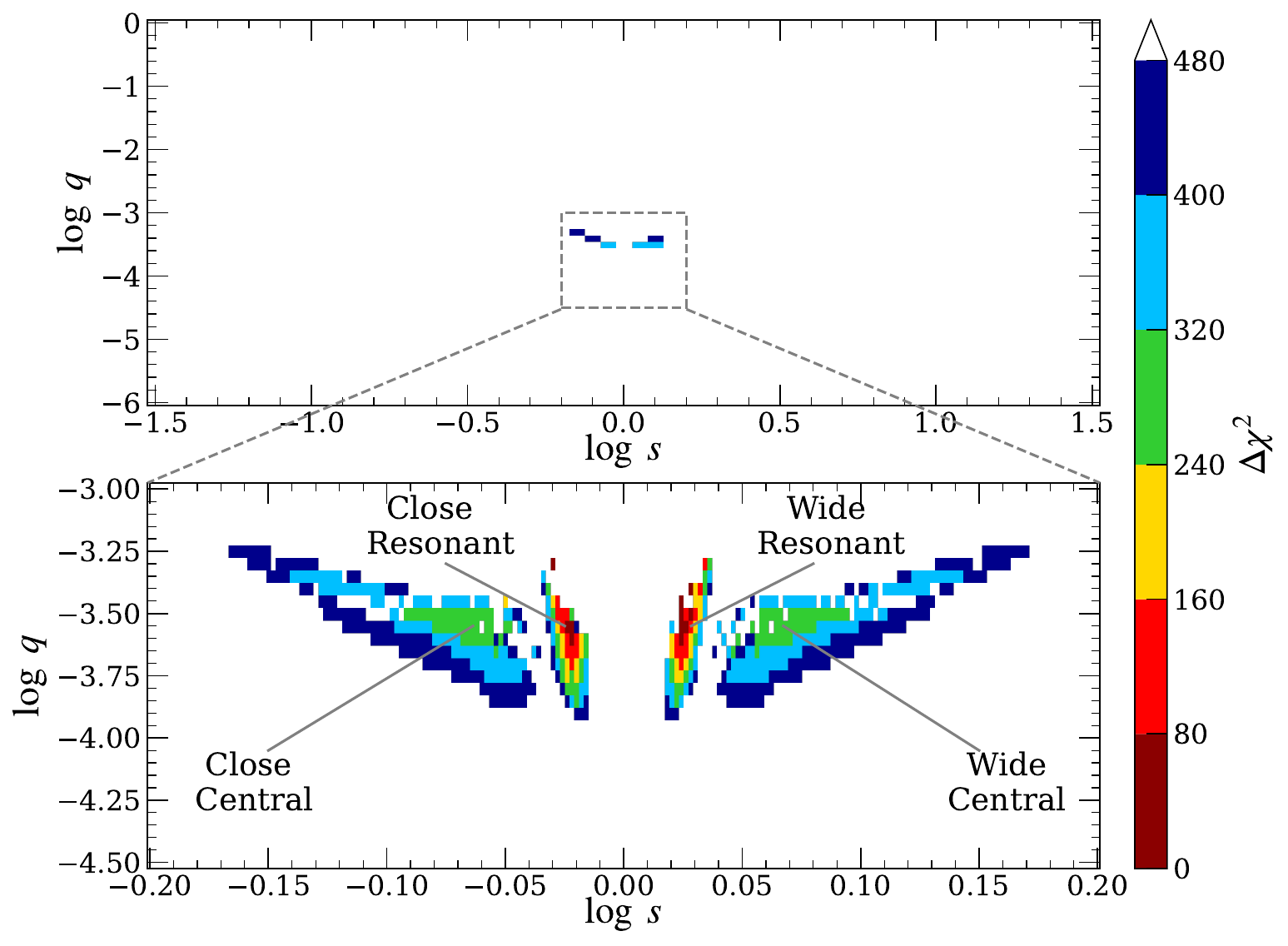}
    \caption{The $\chi^2$ surface in the $(\log s, \log q)$ plane obtained from the 2L1S grid search. The upper panel shows the coarse grid, while the lower panel zooms in on the dense grid around the global minimum. Dark red, red, yellow, green, blue, and dark blue denote grid points within $\Delta\chi^2<1n$, $<2n$, $<3n$, $<4n$, $<5n$, and $<6n$, respectively, where $n = 80$. Grid points with $\Delta\chi^2 > 480$ are left blank. The labels ``Close Central'', ``Wide Central'', ``Close Resonant'', and ``Wide Resonant'' in the lower panel mark the four local minima.}
\label{fig:grid}
\end{figure}

\begin{figure}
    \includegraphics[width=0.47\textwidth]{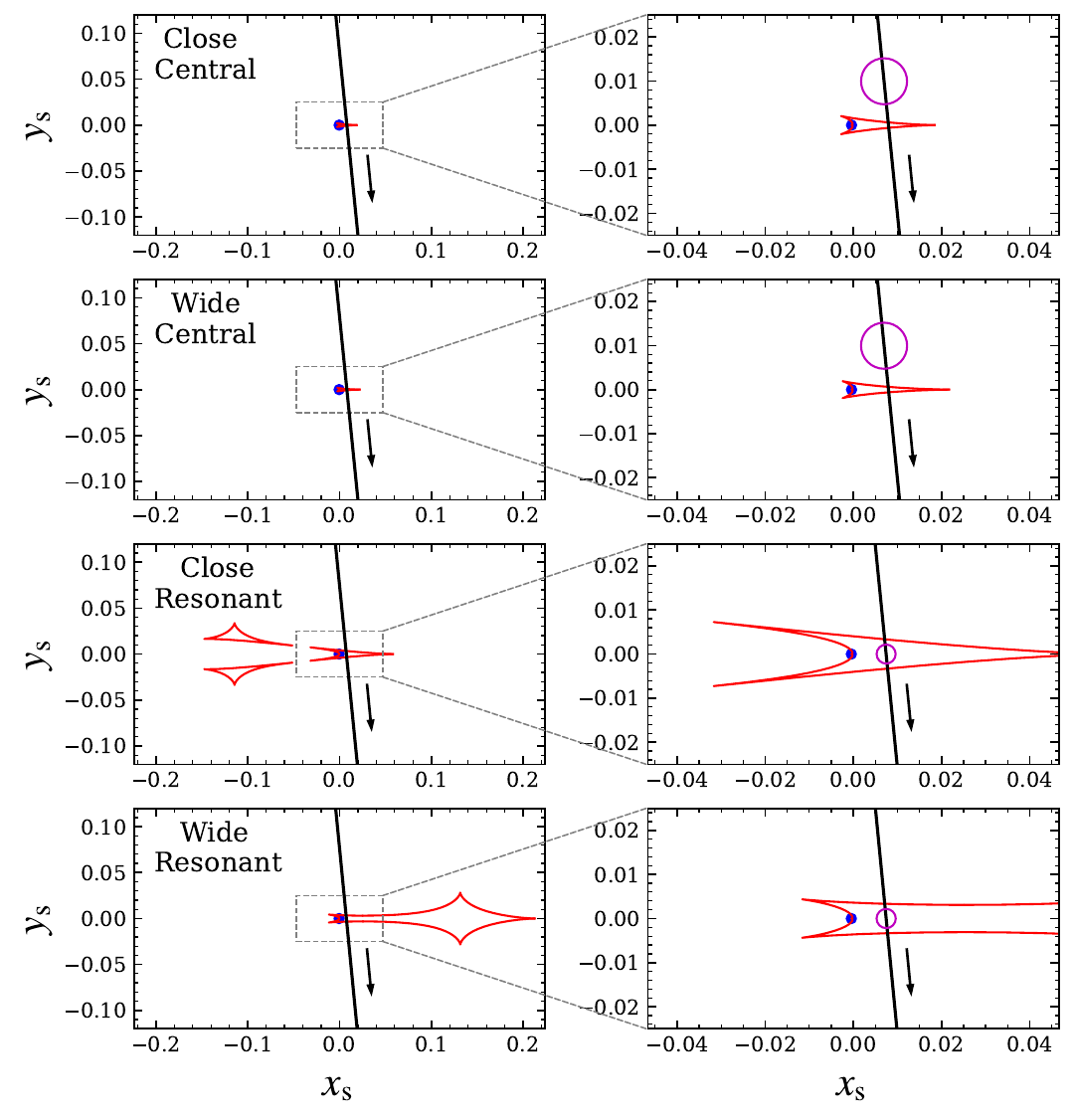}
    \caption{Caustic geometries of the four 2L1S solutions. In each panel, the red lines show the caustic, the blue dot marks the location of the host star, the black line represents the source-lens relative trajectory, and the arrow indicates the direction of source motion. The right panels provide close-up views of the caustic-crossing regions, with the radii of the magenta circles indicating the source size.}
\label{fig:cau}
\end{figure}

Figure \ref{fig:lc} displays all light curves acquired for \event. The source of this event lies at equatorial coordinates of $(\alpha, \delta)_{\rm J2000}$ = (17:51:06.93, $-$30:08:20.80), corresponding to Galactic coordinates of $(\ell,b)$ = ($-0.4137$, $-1.6445)~{\rm deg}$. The event was first detected by the KMTNet AlertFinder \citep{KMTAF} at 02:10~UT on 2025 July 2 (${\rm HJD}^{\prime} = 10858.59, {\rm HJD}^{\prime} = {\rm HJD} - 2450000$). Because KMTNet issued the alert about 20 hours after the event peak, no follow-up observations were obtained from the KMTNet high-magnification follow-up program \citep{KB200414}. The event was independently flagged by the PRIME team 25 hours later as PRIME-2025-BLG-0254. 
 
KMTNet observations were conducted using its three identical 1.6-meter telescopes located at the Siding Spring Observatory (SSO) in Australia (KMTA), CTIO (KMTC), and the South African Astronomical Observatory (SAAO) in South Africa (KMTS). \event\ falls in the overlap of two KMTNet fields, BLG01 and BLG42, giving a combined observing cadence of $4\ \mathrm{hr}^{-1}$; see Figure~12 of \citep{KMTeventfinder} for the field layout. Most KMTNet images used in the light-curve analysis were taken in the $I$ band, while $\sim9\%$ of frames were obtained in the $V$ band to measure the source color. Each KMTNet $V$-band exposure was taken about 2 minutes before or after a KMTNet $I$-band exposure of the same field.

The event is also located in the PRIME GB111 field, which is observed at a cadence of $2.8\ \mathrm{hr}^{-1}$ (see Figure~13 of \citealt{PRIME} for the field placement). PRIME took observations using its 2.2-meter telescope at SAAO. Most PRIME images were taken in the $H$ band, with roughly one $J$-band image per night for source-color measurements. The $J$-band light curve for this event is excluded from the analysis due to insufficient signal-to-noise ratios (SNRs).

In 2025, DREAMS obtained DECam observations on June 29--July 4 and on September 5, 7, and 8. At the position of this event, the June--July run sampled each hour with five $z$-band blocks and one $r$-band block. Each block comprised four exposures, with individual exposure times of 42\,s in $z$ and 60\,s in $r$. Thus, during the peak the DREAMS cadence corresponded to 20 $z$-band and four $r$-band observations per hour. During the September run the cadence was four $z$-band blocks and one $r$-band block per hour. Each $z$-band block contained four images and each $r$-band block three images, with exposure times of 60\,s in $z$ and 80\,s in $r$. Further details of the DREAMS observations will be provided with the first DREAMS data release (Yang, Zang et al., in preparation).

CFHT performed complementary observations with DREAMS on 2025 June 29--30 and July 1 using the $1\ \mathrm{deg}^2$ imager MegaCam. However, the event falls in the gap between two CCDs, so no CFHT data are available for this event.

DECam images were initially processed with the DECam Community Pipeline \citep{DECampipeline}, which performs calibration using reference frames, removes instrumental signatures, and identifies and rejects cosmic rays. The data used in the light-curve analysis were re-reduced with difference imaging analysis (DIA; \citealt{Tomaney1996,Alard1998}) as implemented by each group: \citet{pysis} and \citet{Yang_TLC, Yang_TLC2} for KMTNet and DREAMS, and \citet{Bond2001} for PRIME. The photometric uncertainties produced by the DIA pipelines were renormalized following the procedure of \citet{MB11293}, such that the reduced chi-squared ($\chi^2/{\rm dof}$) for each dataset is unity. The $V$- and $I$-band magnitudes reported in this paper are calibrated to the standard system using the OGLE-III stellar catalog \citep{OGLEIII}, while the $z$- and $r$-band magnitudes are the DECam instrumental magnitude.

%% file: model.tex
As seen in Figure \ref{fig:lc}, the observed light curves of \event\ deviate from the single-lens single-source (1L1S) model (dashed line). A 1L1S model has three Paczy\'{n}ski parameters \citep{Paczynski1986}: $t_0$, $u_0$, and $\te$. These denote, respectively, the time of closest approach of the source to the lens-mass center, the impact parameter in units of the angular Einstein radius $\thetae$, and the Einstein-radius crossing time. The timescale $\te$ is related to the lens mass $M_{\rm L}$, the lens-source relative parallax $\pi_{\rm rel}$, the relative proper motion $\mu_{\rm rel}$, and $\thetae$ via
\begin{equation}\label{eqn:te}
\begin{aligned}
\te &= \frac{\thetae}{\mu_{\rm rel}}; \qquad \thetae = \sqrt{\kappa M_{\rm L} \pi_{\rm rel}};\\
\kappa &\equiv \tfrac{4G}{c^2\,\mathrm{au}} \simeq 8.144~\mathrm{mas}\,M_\odot^{-1}.
\end{aligned}
\end{equation}
The fourth 1L1S parameter is the normalized source size, $\rho\equiv \theta_*/\thetae$, where $\theta_*$ is the angular radius of the source \citep{Shude1994}. 

We fit the light curves with the binary-lens single-source (2L1S, \citealt{Shude1991,Andy1992}) model. A static binary-lens geometry is specified by three parameters: $(q,s,\alpha)$, denoting the binary mass ratio, the projected separation (in units of the Einstein radius), and the angle between the source trajectory and the binary axis, respectively. 

In addition to the seven parameters that determine the magnification over time $f(t)$, for each data set $i$, we include two flux parameters, $(f_{{\rm S},i}, f_{{\rm B},i})$, representing the source flux and any blended flux (from unrelated stars or possible lens light). The observed flux is modeled as
\begin{equation}
    f_i(t) = f_{{\rm S},i}\,A(t) + f_{{\rm B},i},
\end{equation}
where $A(t)$ is the 2L1S magnification as a function of time and computed with the contour-integration code \texttt{VBBinaryLensing} \citep{Bozza2010,Bozza2018,VBMicrolensing2025}.

We incorporate the source surface brightness with a linear limb-darkening law \citep{An2002,Claret2011}. Using the intrinsic source color derived in Section~\ref{lens} together with the color-temperature relations of \citet{Houdashelt2000}, we estimate an effective temperature of $\sim 4830$\,K. Assuming $\log g=2.5$, solar metallicity, and a microturbulent velocity of $1~\mathrm{km\,s^{-1}}$, we adopt the following linear limb-darkening coefficients from \citet{Claret2011}: $u_I=0.57$ ($I$ band), $u_z=0.53$ ($z$ band), $u_r=0.70$ ($r$ band), and $u_H=0.36$ ($H$ band).

To systematically explore the 2L1S parameter space and identify all local $\chi^2$ minima, we adopt a two-stage grid-search strategy. We first run a sparse grid over $(\log s,\log q,\log \rho,\alpha)$ and then refine the solution in the neighborhood of the candidate minima. The coarse grid samples 61 values in $-1.5\le\log s\le1.5$, 61 values in $-6\le\log q\le0$, 9 values in $-4.0\le\log\rho\le-1.6$, and 16 trial angles uniformly in $0\le\alpha<2\pi$. At each grid point we perform a $\chi^2$ minimization with the \texttt{emcee} ensemble sampler \citep{emcee}, holding $(\log s,\log q,\log \rho)$ fixed while allowing $(t_0,u_0,t_{\rm E},\alpha)$ to vary. As shown in the upper panel of Figure 
\ref{fig:grid}, distinct minima are found within $-0.2\le\log s\le0.2$, $-4.5\le\log q\le-3.0$, and $-3.1\le\log\rho\le-1.9$. We therefore carry out a dense follow-up grid comprising 201 values in $-0.2\le\log s\le0.2$, 31 values in $-4.5\le\log q\le-3.0$, 13 values in $-3.1\le\log\rho\le-1.9$, and 16 initial angles that uniformly sample $0\le\alpha<2\pi$.

\begin{table*}[htb]
    \renewcommand\arraystretch{1.20}
    \centering
    \caption{Lensing Parameters using Full Data}
    \begin{tabular}{c|c c c c}
    \hline
    \hline
    \multirow{2}{*}{Parameters} & \multicolumn{2}{c}{Central}  & \multicolumn{2}{c}{Resonant} \\
    & Close & Wide & Close & Wide \\
    \hline
    $\chi^2$/dof & $11762.9/11437$ & $11764.8/11437$ & $11437.5/11437$ & $\mathbf{11433.9/11437}$ \\
    \hline
    $t_{0}-10857$ (${\rm HJD}^{\prime}$)  & $0.76297 \pm 0.00008$ & $0.76298 \pm 0.00008$ & $0.76311 \pm 0.00008$ & $\mathbf{0.76310 \pm 0.00008}$\\
    $u_{0}$ ($10^{-3}$)  & $7.93\pm0.29$ & $7.93\pm0.30$ & $7.40\pm0.28$ & $\mathbf{7.41\pm0.27}$\\
    $\te$ (days)  & $7.91 \pm 0.27$ & $7.91 \pm 0.28$ & $8.05 \pm 0.29$ & $\mathbf{8.03 \pm 0.28}$\\
    $\rho$ ($10^{-3}$) & $5.22 \pm 0.18$ & $5.23 \pm 0.19$ & $2.16 \pm 0.09$ & $\mathbf{2.18 \pm 0.09}$ \\
    $\alpha$ (rad) & $4.8127\pm0.0035$ & $4.8125\pm0.0034$ & $4.8115\pm0.0028$ & $\mathbf{4.8095\pm0.0028}$\\
    $s$ & $0.8637 \pm 0.0050$ & $1.1678 \pm 0.0065$ & $0.9443 \pm 0.0010$ & $\mathbf{1.0678 \pm 0.0012}$\\
    $q (10^{-4})$ & $4.72 \pm 0.24$ & $4.75 \pm 0.26$ & $4.81 \pm 0.22$ & $\mathbf{4.87 \pm 0.23}$\\
    $\log q$ & $-3.325 \pm 0.022$ & $-3.323 \pm 0.024$ & $-3.318 \pm 0.020$ & $\mathbf{-3.312 \pm 0.020}$\\
    $I_{\rm S}$ & $22.036 \pm 0.039$ & $22.037 \pm 0.041$ & $22.054 \pm 0.040$ & $\mathbf{22.054 \pm 0.040}$\\
    \hline
    \hline
    \end{tabular}
    \label{tab:parm1}
\end{table*}

The dense grid search yields four distinct minima, as shown in the lower pabel of Figure \ref{fig:grid}. We then refine each solution with MCMC, allowing all seven 2L1S parameters to vary. Final best-fit models and corresponding $\chi^2$ are obtained via a downhill minimization\footnote{We use the Nelder-Mead simplex implementation in \textsc{SciPy}; see \url{https://docs.scipy.org/doc/scipy/reference/generated/scipy.optimize.fmin.html\#scipy.optimize.fmin}.}. Table~\ref{tab:parm1} lists the corresponding 2L1S parameters with their mean values and $1\sigma$ uncertainties derived from the MCMC posteriors and the $\chi^2$ from the downhill minimization. Figure~\ref{fig:cau} shows the caustic structures and source trajectories for these solutions. 

The four solutions form two close-wide degenerate pairs related approximately by the transformation $s \leftrightarrow s^{-1}$ \citep{Griest1998}, with the remaining parameters nearly unchanged. In one pair, the source crosses the central caustic. Because the source is larger than the caustic width, the resulting light curve shows a single smooth bump over the peak (Figure~\ref{fig:lc}). In the other pair, the source is smaller than the caustic width, producing a ``U''-shaped profile, two caustic crossings connected by a trough. Following the nomenclature of \citet{KB210171,KB220440,KB231431}, we designate these as ``Close Central'' and ``Wide Central'' for the first pair, and ``Close Resonant'' and ``Wide Resonant'' for the second pair. We note that the ``Wide Resonant'' solution exhibits the six-sided resonant caustic, whereas the ``Close Resonant'' solution is not strictly resonant. However, its topology is ``near-resonant'' as proposed by \citet{OB190960}, for which the central and planetary caustics nearly merge, so we retain the resonant designation for this pair.

The ``Wide Resonant'' solution provides the best fit to the light curves, with the ``Close Resonant'' solution only slightly disfavored by $\Delta\chi^2=3.8$. In contrast, the ``Close Central'' and ``Wide Central'' solutions are clearly disfavored by $\Delta\chi^2 \sim 330$. As shown by the residuals in Figure~\ref{fig:lc}, the ``Central'' solutions fail to reproduce the ``U''-shaped profile. We therefore exclude the ``Close Central'' and ``Wide Central'' solutions and retain the ``Close Resonant'' and ``Wide Resonant'' solutions as the viable lensing interpretations of this event. This event provides another example for which the ``Central/Resonant'' degeneracy is resolved by the light curves (e.g., \citealt{KB220440}). The planet-host mass ratio, $q \sim 5\times10^{-4}$, indicates the presence of a planet in the lens system.

Due to the very faint source ($I_{\rm S}\sim22.1$ mag) and short timescale ($\te =8.0\pm0.3$\,days), adding microlensing parallax $\pie$ \citep{Gould1992,Gould2000} improves the fit by only $\Delta\chi^2<0.3$. The parallax components are unconstrained, with $1\sigma$ uncertainties $>1$ in all directions, whereas a typical value is $\sim0.1$. We therefore conclude that this event has no meaningful constraint on parallax.

%% file: lens.tex
\begin{figure}[htb] 
    \centering
    \includegraphics[width=0.47\textwidth]{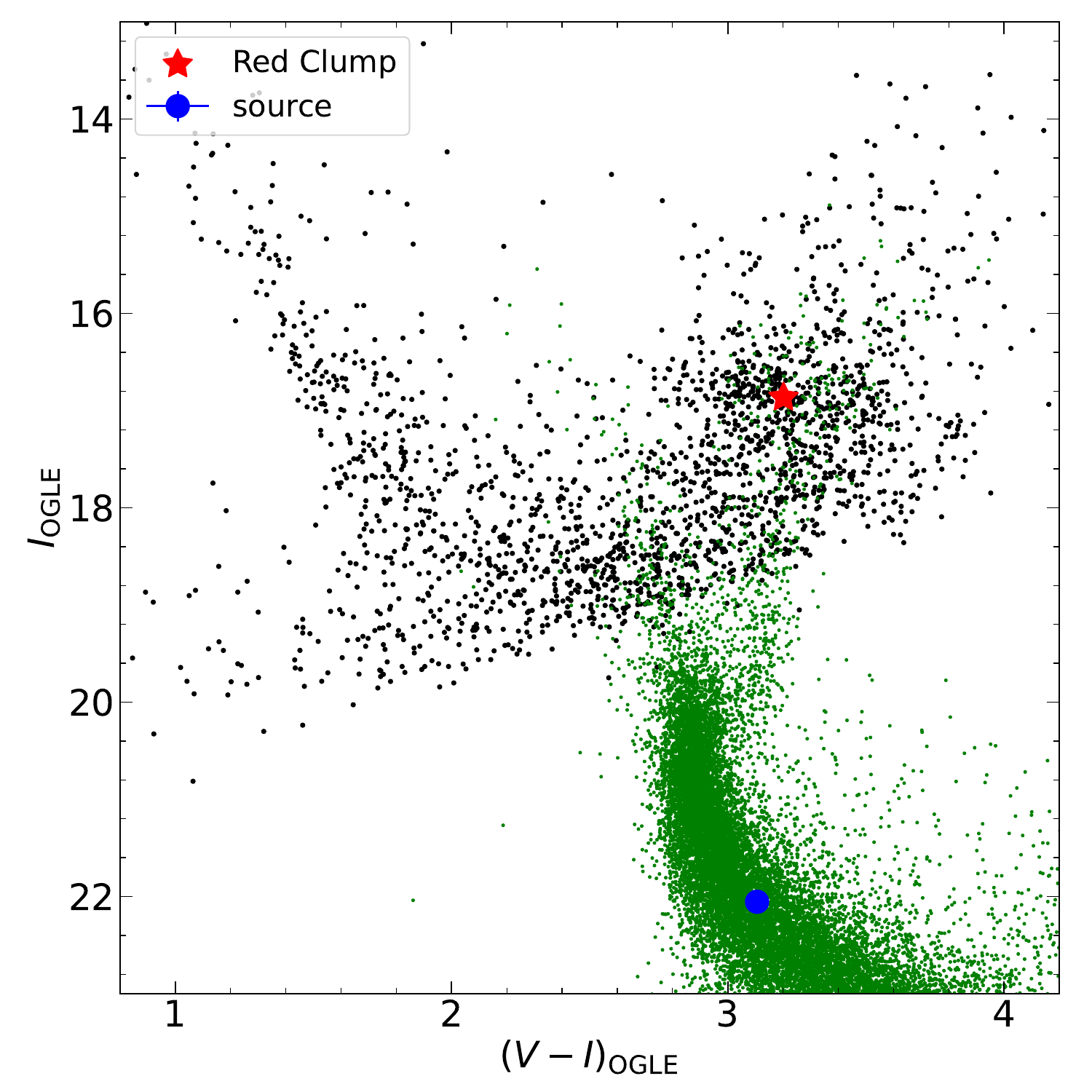}
    \caption{Color-magnitude diagram for \event. Black dots show OGLE-III field stars \citep{OGLEIII} within a $1.5'$ from the event position. The red asterisk and blue dot mark the centroid of the red clump and the source star, respectively. Green points show the \textit{HST} CMD from \cite{HSTCMD}, and the HST red clump centroid, $(V-I,I)_{\rm cl,HST}=(1.62,15.15)$ \citep{MB07192}, has been shifted to match the OGLE-III red clump.}
    \label{fig:cmd}
\end{figure}

\subsection{Color-Magnitude Diagram}\label{CMD}

We estimate the intrinsic color and magnitude of the source by placing it on a color-magnitude diagram (CMD) \citep{Yoo2004}. We construct a $V-I$ versus $I$ CMD from OGLE-III field stars \citep{OGLEIII} within $1.5'$ from the event position. As shown in Figure~\ref{fig:cmd}, the red clump (RC) appears to exhibit an elongated distribution. We find that this apparent elongation is mainly caused by very red stars with magnitudes around $I \sim 17.5$ and colors of $V-I \gtrsim 3.6$, for which the OGLE-III $V$-band photometric uncertainties are relatively large. Removing these stars and using the method of \cite{Nataf2013}, we obtain the centroid of the red clump as $(V - I, I){\rm cl} = (3.202 \pm 0.009, 16.865 \pm 0.016)$. The de-reddened color and magnitude of the red clump are $(V - I, I){\rm cl, 0} = (1.06 \pm 0.03, 14.47 \pm 0.04)$, adopted from \citet{Bensby2013} and \citet{Nataf2013}, so the extinction and reddening toward the event direction are $A_I = 2.395\pm 0.043$ and $E(V - I) = 2.142 \pm 0.031$, respectively.

The source color is first derived via a regression of the KMTC01 $V$ versus $I$ fluxes as a function of magnification and then a calibration to the OGLE-III magnitude system. However, because the event is faint, there is only a single KMTC01 $V$-band measurement (at $\hjd=10857.666$) with a clear magnified signature, yielding $(V-I)_{\rm S}=3.25\pm0.23$, which is too uncertain. There is also one KMTC42 $V$-band point at higher magnification (at $\hjd=10857.778$), but it lies on the anomaly, for which the 2L1S magnifications consider finite-source effects \citep{1994ApJ...421L..75G,Shude1994,Nemiroff1994} and differ between $I$ and $V$ due to band-dependent surface brightness. Even ignoring this effect, we obtain $(V-I)_{\rm S}=3.21\pm0.15$, which remains too uncertain.

We therefore measure the source color using the DREAMS $r-z$ color. Unlike KMTNet's $V$-band protocol, for which an $I$-band exposure is taken within $\sim$2\,min of each $V$-band exposure, DREAMS acquires $r$ and $z$ data in separate blocks, with a typical gap of $\sim$10\,min between an $r$ block and the nearest $z$ block. Accordingly, we first bin each $r$-band block to a single point and then bin all $z$-band measurements taken within 15\,min of that binned $r$ point. The resulting pairs of binned $r$ and $z$ points are used in a linear regression to derive the source color. The mean $r$-$z$ time offset is 2.7\,min.

We exclude three DREAMS $r$-band blocks obtained during finite-source effects at $\hjd=10857.72,\,10857.76,$ and $10857.81$. Our procedure yields $(r-z)_{\rm S, DECam}=1.212\pm0.006$, consistent with the MCMC posterior mean, $(r-z)_{\rm S, DECam}=1.200$, from the 2L1S modeling. Calibrating with bright field stars common to the OGLE-III and DECam images yields $(V-I)_{\rm S}=3.106\pm0.016$, which is sufficiently precise. The source's intrinsic color and magnitude are then $(V-I, I)_{\rm S,0} = (0.96 \pm 0.04,\ 19.66 \pm 0.06)$. In Figure~\ref{fig:cmd}, we also align the {\it HST} CMD from \citet{HSTCMD} with the OGLE-III CMD. This comparison shows that the source color derived from the DREAMS $r-z$ color is typical for a star with $I_0=19.66$.

Applying the color-surface-brightness relation of \citet{Adams2018}, we derive an angular source radius of $\theta_* = 0.470 \pm 0.026~\mu{\rm as}$. This implies
\begin{numcases}{\thetae=\frac{\theta_*}{\rho}=}
0.217 \pm 0.014~\mathrm{mas} & {\rm for\ Close\ Resonant,} \\
0.216 \pm 0.014~\mathrm{mas} & {\rm for\ Wide\ Resonant.}
\end{numcases}
and the geocentric lens-source relative proper motion
\begin{numcases}{\mu_{\rm rel}=\frac{\thetae}{\te}=}
9.85 \pm 0.74~\mathrm{mas\,yr^{-1}} & {\rm for\ Close\ Resonant,} \\
9.82 \pm 0.74~\mathrm{mas\,yr^{-1}} & {\rm for\ Wide\ Resonant.}
\end{numcases}

\subsection{Bayesian Analysis}\label{Baye}

\begin{table*}[htb]
    \renewcommand\arraystretch{1.25}
    \centering
    \caption{Lensing Physical Parameters from a Bayesian Analysis}
    \begin{tabular}{c |c c c c c}
    \hline
    \hline
     \multirow{2}{*}{Model} &\multicolumn{5}{c}{Physical Properties} \\
     & $M_\mathrm{host}(M_\odot)$ & $M_\mathrm{planet}(M_\oplus)$ & $D_\mathrm{L}(\mathrm{kpc})$ & $a_\bot(\mathrm{au})$ & $\mu_\mathrm{rel, hel}(\mathrm{mas\ yr^{-1}})$  \\
    \hline
    Close Resonant  & $0.27_{-0.14}^{+0.25}$ & $42.30_{-22.03}^{+40.24}$&$7.5_{-1.0}^{+0.7}$&$1.5_{-0.2}^{+0.2}$&$9.53_{-0.70}^{+0.73}$\\
    Wide Resonant & $0.26_{-0.14}^{+0.25}$ & $42.32_{-22.08}^{+40.70}$&$7.5_{-1.0}^{+0.7}$&$1.7_{-0.2}^{+0.2}$&$9.47_{-0.70}^{+0.73}$ \\
    Combined & $0.26_{-0.14}^{+0.25}$ & $42.32_{-22.07}^{+40.62}$&$7.5_{-1.0}^{+0.7}$&$1.6_{-0.3}^{+0.2}$&$9.48_{-0.70}^{+0.73}$ \\
    \hline
    \hline
    \end{tabular}
     \tablecomments{The combined model is obtained by a combination of two models weighted by the probability for the Galactic model and ${\rm exp}(-\Delta\chi^2/2)$.}
    \label{tab:baye}
\end{table*}

\begin{figure*}[htb] 
    \centering
    \includegraphics[width=0.9\textwidth]{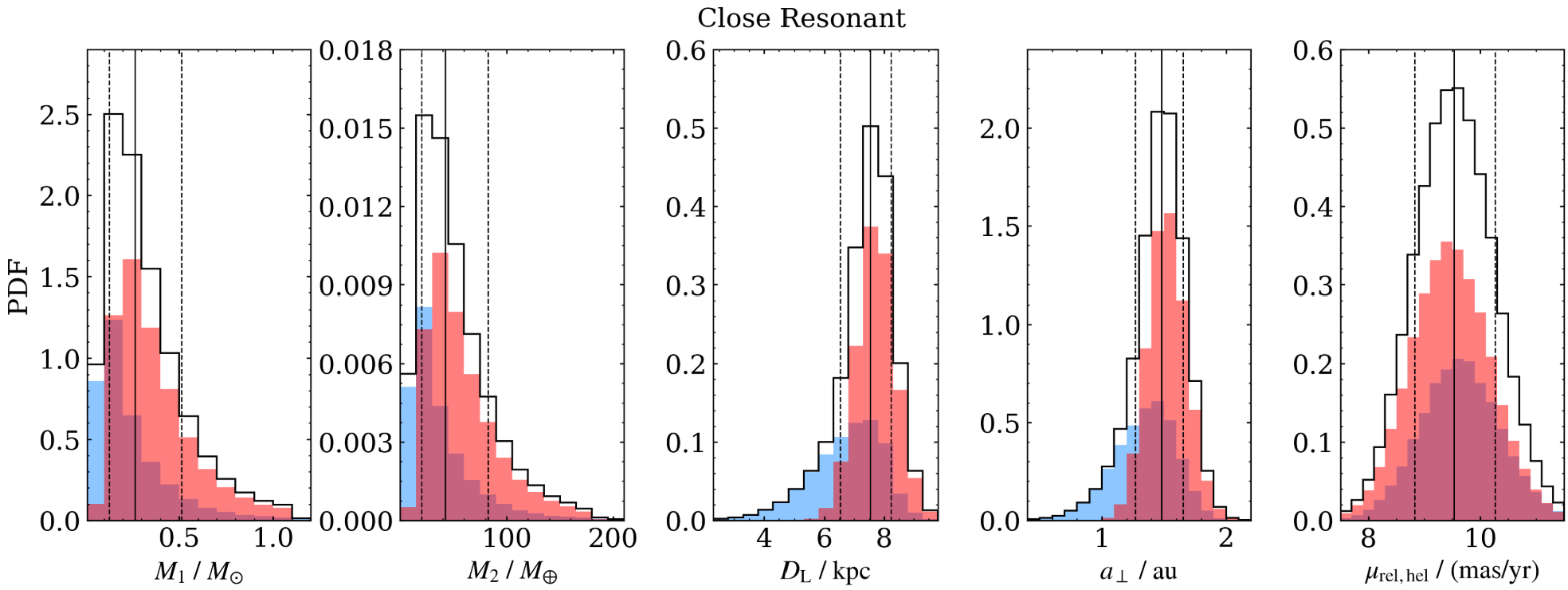}
    \includegraphics[width=0.9\textwidth]{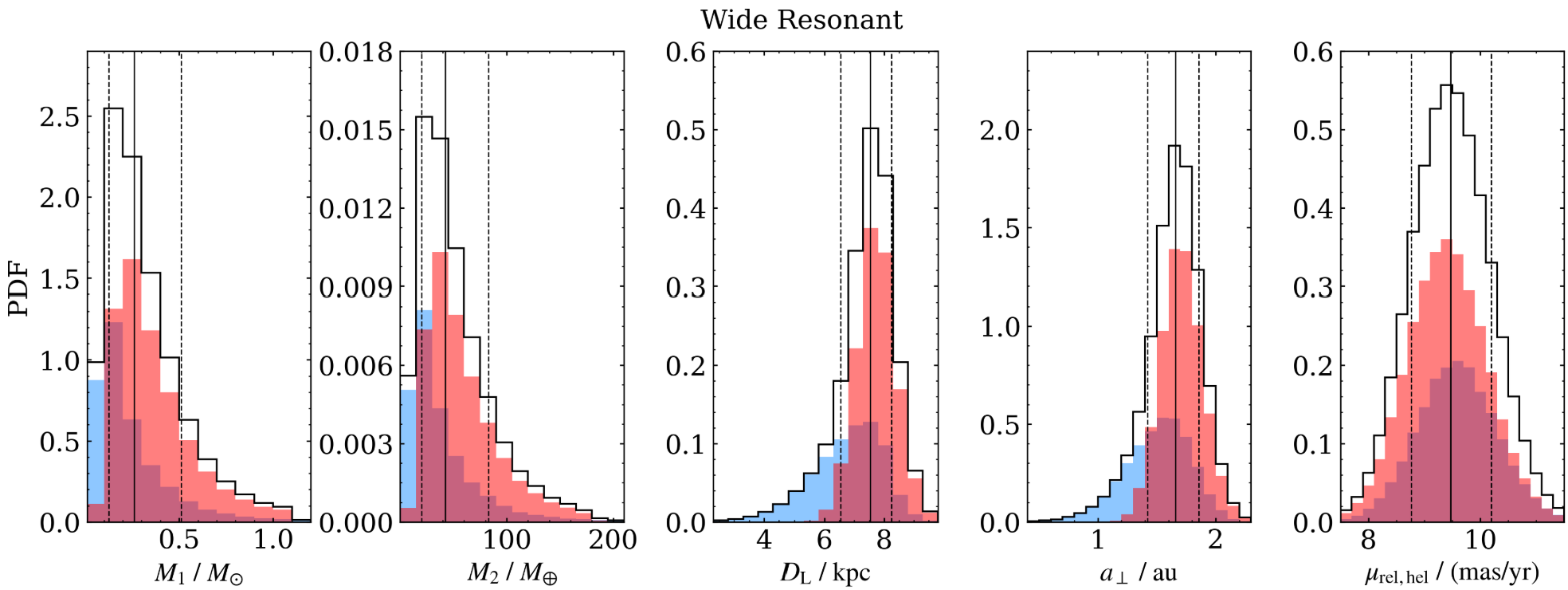}
    \caption{Posterior probability distributions from our Bayesian analysis are shown for the host mass $M_1$, the planetary mass $M_2$, the lens distance $D_{\rm L}$, the projected planet-host separation $a_\bot$, and the heliocentric lens-source relative proper motion $\mu_{\rm hel,rel}$. In each panel, the solid black curve indicates the median, and the two dashed black curves mark the 15.9\% and 84.1\% credible limits. Contributions from bulge and disk lens populations are shown in red and blue, respectively.}
    \label{fig:baye}
\end{figure*}

The lens mass $M_{\rm L}$ and distance $D_{\rm L}$ are related to the angular Einstein radius and the microlensing parallax by \citep{Gould1992,Gould2000}
\begin{equation}\label{eq:mass}
    M_{\rm L} = \frac{\theta_{\rm E}}{\kappa \pi_{\rm E}}, \qquad 
    D_{\rm L} = \frac{\mathrm{au}}{\pi_{\rm E}\theta_{\rm E} + \mathrm{au}/D_{\rm S}},
\end{equation}
where $D_{\rm S}$ is the source distance. Because this event provides no meaningful constraint on $\pi_{\rm E}$, we estimate the physical parameters of the lens system through a Bayesian analysis, adopting priors from a Galactic model.

Our Galactic prior comprises three components: the lens mass function, the spatial number density distributions of lenses and sources, and their kinematics. For the lens mass function, we adopt the \citet{Kroupa2001} initial mass function, imposing upper-mass cutoffs of $1.3\,M_\odot$ for disk lenses and $1.1\,M_\odot$ for bulge lenses \citep{Zhu2017spitzer}. The stellar density profiles for both the disk and the bulge follow the prescriptions in \citet{Yang2021_GalacticModel}. For the bulge velocity distribution, we use the same framework as \citet{Zhu2017spitzer}, while the disk kinematics follow the ``Model '' of \citet{Yang2021_GalacticModel}, implemented with the \texttt{galpy} package \citep{bovy2015galpy}.

We simulate a sample of $10^8$ events. For each event $i$, characterized by $(t_{{\rm E},i},\,\mu_{{\rm rel},i},\,\theta_{{\rm E},i})$, we assign a weight
\begin{equation}\label{equ: weight}
    w_{i} = \Gamma_{i}\times p(t_{{\rm E}, i}) p(\theta_{{\rm E},i}),
\end{equation}
where $\Gamma_i \equiv \theta_{{\rm E},i}\,\mu_{{\rm rel},i}$ is the microlensing event rate, and $p(t_{{\rm E},i})$ and $p(\theta_{{\rm E},i})$ are the likelihood factors for $t_{{\rm E},i}$ and $\theta_{{\rm E},i}$ given the probability distributions for each solution.

The physical parameters inferred from the Bayesian analysis are summarized in Table~\ref{tab:baye} and illustrated in Figure~\ref{fig:baye}. These include the host mass $M_1$, the companion (planetary) mass $M_2$, the lens distance $D_{\rm L}$, the projected star-planet separation $a_\perp$, and the heliocentric lens-source relative proper motion $\mu_{\rm rel, hel}$. The results favor an M-dwarf host located in the Galactic bulge. The companion mass lies between that of Neptune and Saturn, with a projected separation of about 1.6\,au. Using a water snow-line scaling of $a_{\rm SL}=2.7(M/M_\odot)$\,au \citep{snowline}, the planet is situated well beyond the snow line.

%% file: dis.tex
\subsection{DREAMS's Role in This Planetary Detection}

\begin{table*}[htb]
    \renewcommand\arraystretch{1.20}
    \centering
    \caption{Lensing Parameters without the DREAMS Data}
    \begin{tabular}{c|c|c c c c}
    \hline
    \hline
    \multirow{3}{*}{Parameters} & \multirow{3}{*}{1L1S} & \multicolumn{4}{c}{2L1S} \\
    & & \multicolumn{2}{c}{Central}  & \multicolumn{2}{c}{Resonant} \\
    & & Close & Wide & Close & Wide \\
    \hline
    $\chi^2$/dof & $10487.0/10414$ & $\mathbf{10399.5/10411}$ & $10399.6/10411$ & $11400.0/10411$ & $11400.0/10411$ \\
    \hline
    $t_{0}-10857$ (${\rm HJD}^{\prime}$) & $76423 \pm 0.00033$ & $\mathbf{0.76311 \pm 0.00046}$ & $0.76316 \pm 0.00046$ & $0.76333 \pm 0.00043$ & $0.76329 \pm 0.00044$\\
    $u_{0}$ ($10^{-3}$) &  $7.38 \pm 0.58$ & $\mathbf{7.99\pm0.71}$ & $7.93\pm0.69$ & $7.63\pm0.71$ & $7.61\pm0.66$\\
    $\te$ (days) & $9.46 \pm 0.71$ & $\mathbf{8.08 \pm 0.63}$ & $8.12 \pm 0.62$ & $8.10 \pm 0.64$ & $8.11 \pm 0.60$\\
    $\rho$ ($10^{-3}$) & $7.83 \pm 0.63$ & $\mathbf{4.92 \pm 0.42}$ & $4.89 \pm 0.42$ & $2.36 \pm 0.33$ & $2.37 \pm 0.32$ \\
    $\alpha$ (rad) & ... & $\mathbf{4.8025\pm0.0185}$ & $4.8011\pm0.0179$ & $4.8007\pm0.0186$ & $4.8003\pm0.0184$\\
    $s$ & ... & $\mathbf{0.8336 \pm 0.0499}$ & $1.2056 \pm 0.0665$ & $0.9438 \pm 0.0063$ & $1.0684 \pm 0.0076$\\
    $q (10^{-4})$ & ... & $\mathbf{4.81 \pm 1.11}$ & $4.65 \pm 1.04$ & $4.60 \pm 0.82$ & $4.64 \pm 0.81$\\
    $\log q$ & ... & $\mathbf{-3.318 \pm 0.100}$ & $-3.332 \pm 0.097$ & $-3.337 \pm 0.077$ & $-3.334 \pm 0.076$\\
    $I_{\rm S}$ & $22.282 \pm 0.085$ & $\mathbf{22.011 \pm 0.092}$ & $21.991 \pm 0.088$ & $22.020 \pm 0.095$ & $22.025 \pm 0.090$\\
    \hline
    \hline
    \end{tabular}
    \label{tab:parm2}
\end{table*}

\begin{figure*}
    \centering
    \includegraphics[width=0.85\linewidth]{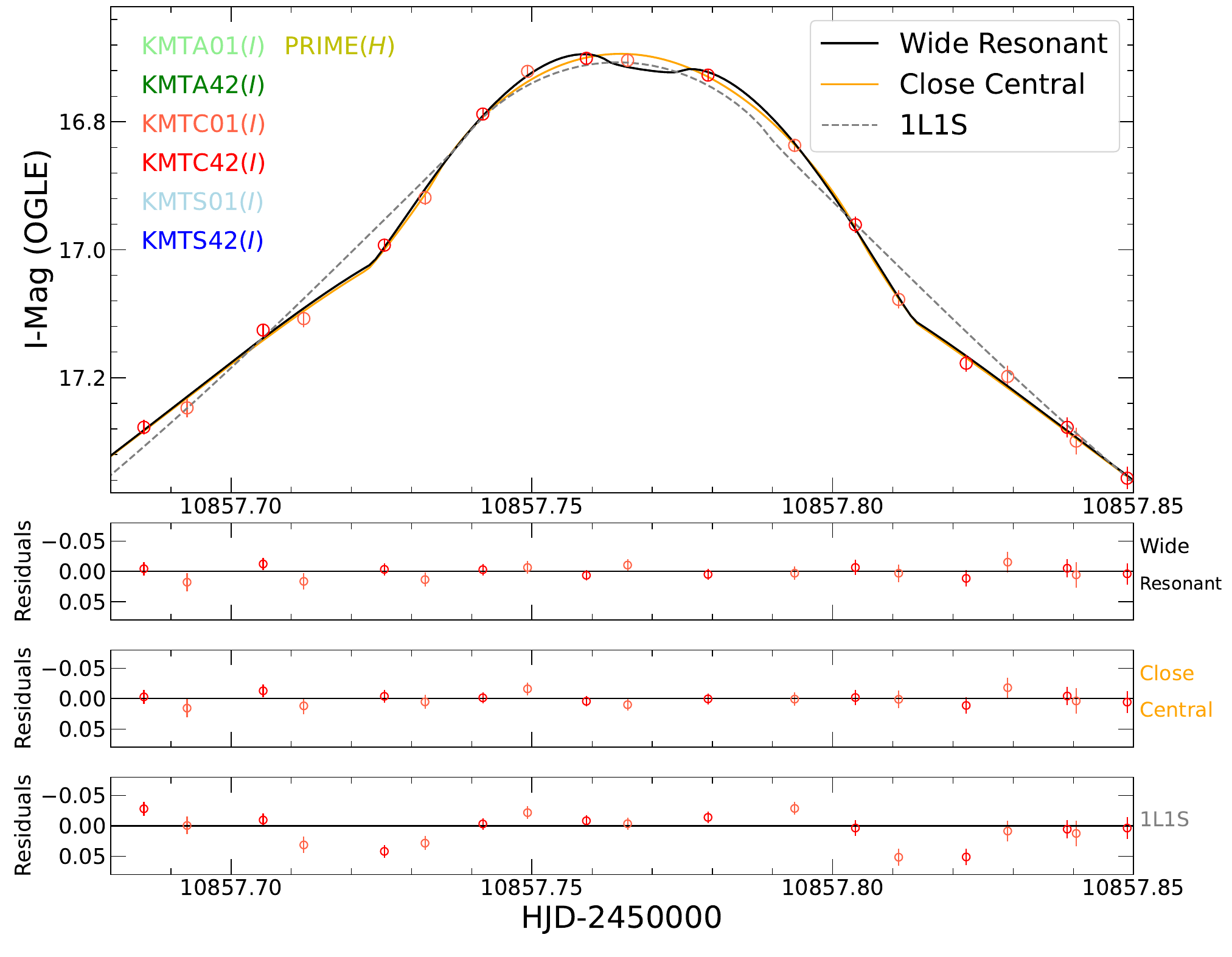}
    \caption{A close-up of the anomaly and models excluding the DREAMS data. The symbols are similar to those in Figure \ref{fig:lc}. The parameters are shown in Table \ref{tab:parm2}. }
    \label{fig:lc2}
\end{figure*}

The anomaly of \event\ was identified in two independent ways. C. Han detected it using KMTNet's online pySIS photometry, while the DREAMS team independently discovered it by examining the DREAMS light curves of this high-magnification event. We find that the anomaly from the online photometry can be identified by applying the KMTNet AnomalyFinder algorithm \citep{OB191053,2019_prime}. 

We further perform 1L1S and 2L1S modeling without the DREAMS data to check the role of the data (other datasets are identical used in Section \ref{model}). Table~\ref{tab:parm2} summarizes the parameters, and Figure~\ref{fig:lc2} shows the corresponding model light curves and residuals. Without the DREAMS data, the light curves also yield two pairs of degenerate 2L1S solutions: ``Close Central'', ``Wide Central'', ``Close Resonant'', and ``Wide Resonant''. Their parameters are consistent with those obtained including the DREAMS data (Table~\ref{tab:parm1}) within $1\sigma$, although the uncertainties in the binary-lens geometry ($q$, $s$, $\alpha$) and $\rho$ increase by factors of 3--6. Crucially, in the absence of DREAMS data, the ``Central/Resonant'' degeneracy is severe, with $\Delta\chi^2 = 0.5$ among the four solutions .\cite{KB210171} suggested that the $(\log s, \log q, \alpha)$ phase-space factors can be used to weight the probability of each degenerate solution. For this event, the phase-space factors of the central solutions are approximately 10 times larger than those of the resonant solutions. Therefore, if the DREAMS data were unavailable and the degeneracy were weighted  by the phase-space factors, the ``Central'' solutions would be incorrectly favored.

Therefore, the DREAMS data provide three key contributions to this event. First, they break the ``Central/Resonant'' degeneracy. Second, they substantially tighten the binary-lens parameters. Third, as shown in Section~\ref{CMD}, the event is too faint for KMTNet $V$-band measurements, and DREAMS's high-cadence $r$-band observations yield a precise source color. The DREAMS $r-z$ color can be used to constrain source colors in other faint events, even when DREAMS does not cover the planetary anomaly. Moreover, for faint events in high-extinction fields, the combined $I - z$ color from KMTNet and DREAMS could also constrain the source color.

\subsection{An Estimate of DREAMS's Detection Limits for FFPs}

\begin{figure*}
    \centering
    \includegraphics[width=0.85\linewidth]{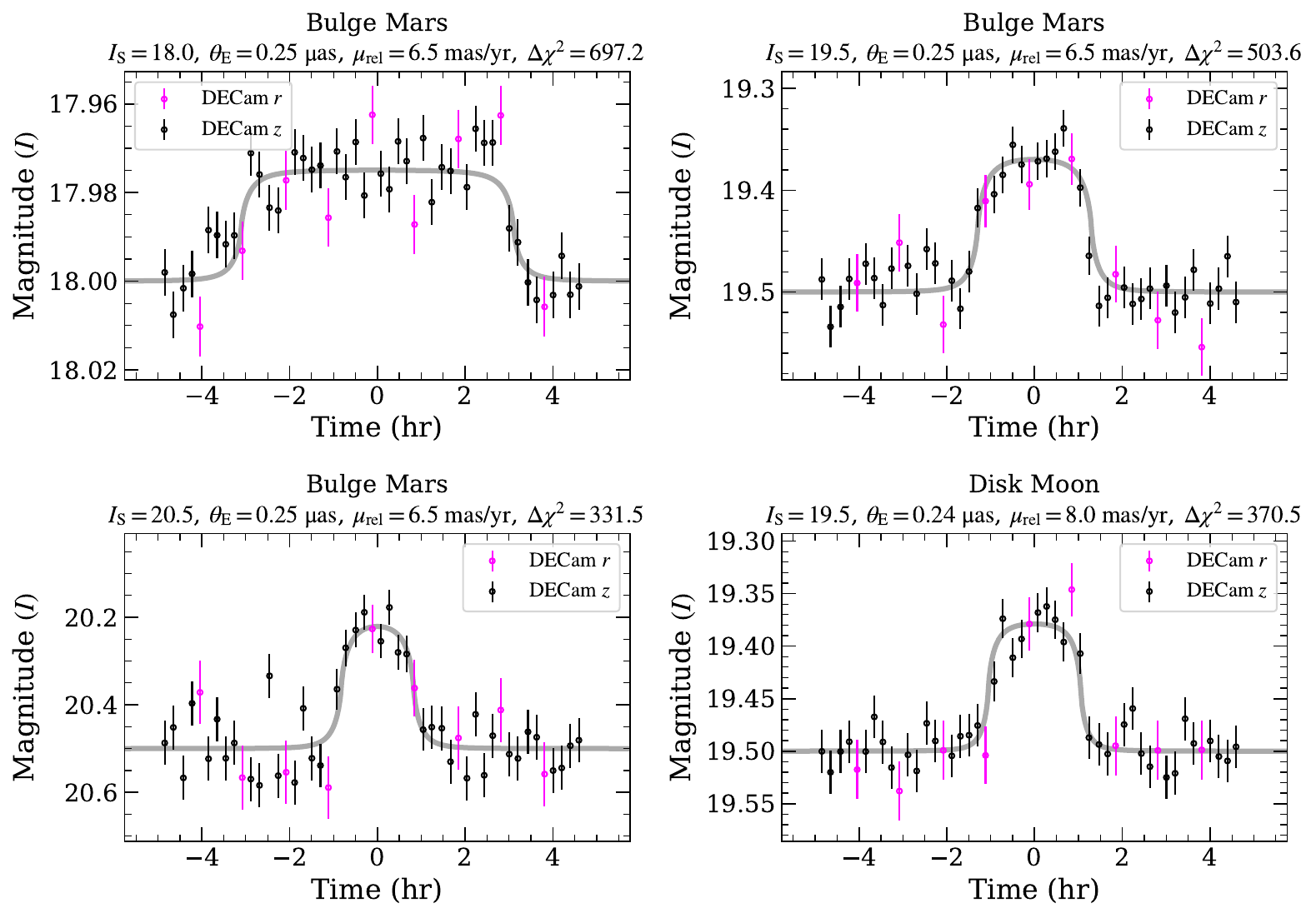}
    \caption{Simulated FFP events as observed by DREAMS. For clarity, data within each observing block are binned (four points in $z$ and three in $r$ per block). Magnitudes are placed on the $I$-band scale. Lens and source for each panel parameters are summarized in Table~\ref{tab:simu}. $\Delta\chi^2$ denotes the difference in $\chi^2$ between a flat model and the best-fit 1L1S model.}
    \label{fig:ffp}
\end{figure*}

The primary motivation of DREAMS is to detect low-mass FFPs with minute-level cadence. Based on the practical photometric accuracy of the DREAMS observations obtained from this event, we can estimate the minimum FFP masses that DREAMS can detect.

The current plan for DREAMS observations in 2026--2028 will adopt the same survey strategy as the September 2025 run. Specifically, the survey will cover two fields with a $1~\mathrm{deg}^2$ overlap. Each field will be observed with four $z$-band blocks and one $r$-band block per hour, with exposure times of $60\,\mathrm{s}$ in $z$ and $80\,\mathrm{s}$ in $r$. For estimating the detection limits, we adopt the strategy corresponding to a single field.

The photometric SNRs are taken from the standard deviations of the September 2025 run, with SNR = 26.5 at $z_{\rm DECam} = 18.4$ mag and SNR = 14.7 at $r_{\rm DECam} = 19.4$ mag. For the simulated events we show below, the noise is dominated by sky background from airglow and moonlight, since DREAMS operates during DECam bright time. Consequently, despite variations in extinction and stellar surface density across the DREAMS fields, these photometric error bars remain applicable. The SNRs for different brightness are estimated by 
\begin{align}
    {\rm SNR}(z_{\rm DECam}) &= 26.5 \times 10^{(18.4 - z_{\rm DECam})/2.5}, \\
    {\rm SNR}(r_{\rm DECam}) &= 14.7 \times 10^{(19.4 - z_{\rm DECam})/2.5}.
\end{align}

The DREAMS survey fields are located at $-1.0 \lesssim \ell \lesssim 2.0$ and $-2.8 \lesssim b \lesssim -0.6$, with a central position at approximately $(\ell, b) = (0.5, -1.8)$. We adopt a representative extinction of $A_I = 1.8$ and a reddening of $E(V-I) = 1.55$, which are close to the extinction values at the field center \citep{Nataf2013} and are lower than those of the field surrounding \event. To transform DECam instrumental magnitudes to the standard $V$ and $I$ system commonly used, we match bright field stars in common between DREAMS and OGLE-III and obtain
\begin{align}
    z_{\rm DECam} &= I - 0.55 - 0.24\,(V - I), \\
    r_{\rm DECam} &= I - 1.75 + 0.53\,(V - I).
\end{align}

We first consider DREAMS's sensitivity to Mars-mass FFPs in the Galactic bulge. We adopt a typical bulge configuration with $D_{\rm S}=8.5~\mathrm{kpc}$, $D_{\rm L}=7.0~\mathrm{kpc}$, and $\mu_{\rm rel}=6.5~\mathrm{mas\,yr^{-1}}$ (The proper motion value is adopted from \citealt{CMST}). We simulate three cases with source brightnesses $I=18.0$, $19.5$, and $20.5$, respectively; their colors are assigned as the median color of {\it HST} field stars whose extinction-corrected magnitudes are within $0.1$~mag of the corresponding source.

Figure~\ref{fig:ffp} displays the simulated light curves. We bin the data within each cadence group into a single point. All cases satisfy the FFP detection criteria adopted in the {\it Roman} FFP simulations \citep{Johnson2020}: (1) at least six data points exceed the baseline flux by $\ge 3\sigma$; and (2) the $\chi^2$ difference between a flat model and the best-fit 1L1S model is $\ge 300$. The $I=20.5$ case yields the smallest $\Delta\chi^2$ and lies closest to the detection threshold, implying that, for a typical DREAMS field, sources with $I \lesssim 20.5$ are sensitive to Mars-mass FFPs. For the $I=18$ source, the $z$-band points rise to $\sim 4\sigma$ above baseline for a Mars-mass FFP. Scaling to a $3\sigma$ criterion therefore reduces the corresponding mass to $3/4~M_{\rm Mars}$, and the resulting $\Delta\chi^2 \sim 390$ remains above the detection limit. A comparable threshold mass is obtained for the $I=19.5$ source, for which the $\Delta\chi^2$ requirement sets the detection limit.

Because disk lenses have larger $\thetae$, Mars-mass FFPs in the disk yield higher SNRs. We therefore investigate whether Moon-mass FFPs can be detected by DREAMS. The lower right panel of Figure~\ref{fig:ffp} shows a simulated light curve for a Moon-mass FFP with an $I=19.5$ source and $D_{\rm L}=3.2~\mathrm{kpc}$. We adopt a slightly larger lens-source relative proper motion, $\mu_{\rm rel}=8~\mathrm{mas\,yr^{-1}}$. The simulated light curve contains 10 data points that exceed the baseline flux by $\ge 3\sigma$ and yields $\Delta\chi^{2}=370$, indicating that DREAMS is sensitive to disk Moon-mass FFPs. In the $1~\mathrm{deg}^2$ region where the two DREAMS fields overlap, the cadence is double and thus DREAMS is sensitive to planets less massive than the Moon. Lower extinction increases the source flux received by DECam and thus enables sensitivity to less-massive FFPs; conversely, higher extinction reduces the sensitivity. We have been developing the DREAMS full-frame imaging pipeline following the framework of \cite{KMTFFP1}, and we expect to complete the FFP search for 2025 observations by early 2026.

\begin{table}[htb]
    \renewcommand\arraystretch{1.25}
    \centering
    \caption{Lens and Source Information for Simulated Events}
    \begin{tabular}{c |c c c|c}
    \hline
    \hline
     Parameters & \multicolumn{3}{c|}{Bulge Mars} & Disk Moon \\
     \hline
     $I_{\rm S}$ & 18.00 & 19.50 & 20.50 & 19.50 \\
     $V_{\rm S}$ & 20.50 & 21.80 & 22.80 & 21.80 \\
     $z_{\rm DECam}$ & 16.85 & 18.40 & 19.40 & 18.40 \\
     $r_{\rm DECam}$ & 17.55 & 18.95 & 19.95 & 18.95 \\
     $D_{\rm S}$ (kpc) & 8.5 & 8.5 & 8.5 & 8.5 \\
     $\theta_*$ ($\mu$as) & 2.3 & 0.96 & 0.61 & 0.96 \\
     $M_{\rm L}$ ($M_{\oplus}$) & 0.107 & 0.107 & 0.107 & 0.0123 \\
     $D_{\rm L}$ (kpc) & 7.0 & 7.0 & 7.0 & 3.2 \\
     $\thetae$ ($\mu$as) & 0.25 & 0.25 & 0.25 & 0.24 \\
     $\mu_{\rm rel}$ ($\mathrm{mas\,yr^{-1}}$) & 6.5 & 6.5 & 6.5 & 8.0 \\
    \hline
    \hline
    \end{tabular}
     \tablecomments{Assume $A_I = 1.8$ and $E(V-I) = 1.55$.}
    \label{tab:simu}
\end{table}

\acknowledgments
H.Y. acknowledge support by the China Postdoctoral Science Foundation (No. 2024M762938). W.Z. acknowledges the support from the Harvard-Smithsonian Center for Astrophysics through the CfA Fellowship. H.Y., W.Z., J.Z., H.L., Y.T., Q.Q., Z.L., Y.S., X.S. and S.M. acknowledge support by the National Natural Science Foundation of China (Grant No. 12133005). This work is part of the ET space mission which is funded by the China's Space Origins Exploration Program. This research has made use of the KMTNet system operated by the Korea Astronomy and Space Science Institute (KASI) at three host sites of CTIO in Chile, SAAO in South Africa, and SSO in Australia. Data transfer from the host site to KASI was supported by the Korea Research Environment Open NETwork (KREONET). This research was supported by KASI under the R\&D program (project No. 2025-1-830-05) supervised by the Ministry of Science and ICT. The PRIME project is supported by JSPS KAKENHI Grant Number JP16H06287, JP22H00153, JP25H00668,
JP19KK0082, JP20H04754, JP24H01811 and JPJSCCA20210003. The PRIME project acknowledges a financial support by Astrobiology Center. Work by J.C.Y. acknowledge support from N.S.F Grant No. AST-2108414. Work by C.H. was supported by the grants of National Research Foundation of Korea (2019R1A2C2085965 and 2020R1A4A2002885). Y.S. acknowledges support from BSF Grant No. 2020740. RAS and KK acknowledge support from US National Science Foundation grant 2206828.

This project used data obtained with the Dark Energy Camera (DECam), which was constructed by the Dark Energy Survey (DES) collaboration. Funding for the DES Projects has been provided by the U.S. Department of Energy, the U.S. National Science Foundation, the Ministry of Science and Education of Spain, the Science and Technology Facilities Council of the United Kingdom, the Higher Education Funding Council for England, the National Center for Supercomputing Applications at the University of Illinois at Urbana-Champaign, the Kavli Institute for Cosmological Physics at the University of Chicago, the Center for Cosmology and Astro-Particle Physics at The Ohio State University, the Mitchell Institute for Fundamental Physics and Astronomy at Texas A\&M University, Financiadora de Estudos e Projetos, Funda\c{c}\~ao Carlos Chagas Filho de Amparo \`a Pesquisa do Estado do Rio de Janeiro, Conselho Nacional de Desenvolvimento Cient\'{\i}fico e Tecnol\'ogico and the Minist\'erio da Ci\^encia, Tecnologia e Inova\c{c}\~ao, the Deutsche Forschungsgemeinschaft, and the collaborating institutions in the Dark Energy Survey.

The collaborating institutions are Argonne National Laboratory; the University of California at Santa Cruz; the University of Cambridge; Centro de Investigaciones Energ\'eticas, Medioambientales y Tecnol\'ogicas (CIEMAT), Madrid; the University of Chicago; University College London; the DES-Brazil Consortium; the University of Edinburgh; the Eidgen\"ossische Technische Hochschule (ETH) Z\"urich; Fermi National Accelerator Laboratory; the University of Illinois at Urbana-Champaign; the Institut de Ci\`encies de l'Espai (IEEC/CSIC); the Institut de F\'{\i}sica d'Altes Energies (IFAE); Lawrence Berkeley National Laboratory; the Ludwig-Maximilians-Universit\"at M\"unchen and the associated Excellence Cluster Universe; the University of Michigan; NSF NOIRLab; the University of Nottingham; The Ohio State University; the OzDES Membership Consortium; the University of Pennsylvania; the University of Portsmouth; SLAC National Accelerator Laboratory; Stanford University; the University of Sussex; and Texas A\&M University.

Based on observations at NSF Cerro Tololo Inter-American Observatory, NSF NOIRLab (NOIRLab Prop.\ ID 2025A-806294, PI: Weicheng Zang; Prop.\ ID 2025B-560332, PI: Weicheng Zang \& Hongjing Yang), which is managed by the Association of Universities for Research in Astronomy (AURA) under a cooperative agreement with the U.S. National Science Foundation.

\software{pySIS \citep{pysis,Yang_TLC,Yang_TLC2}, numpy \citep{numpy}, emcee \citep{emcee2,emcee}, Matplotlib \citep{Matplotlib}, SciPy \citep{scipy}, galpy \citep{bovy2015galpy}, DECam Community Pipeline \citep{DECampipeline}}